\documentstyle[rep12]{report}

\newtheorem{FIG}{Fig.}
\def\figure#1{\FIG{\rm{#1}}}

\title{
Photoinduced charge separation in Q1D heterojunction materials:\\
Evidence for electron-hole pair separation in mixed-halide $MX$
solids
}
\author{
L.A.~Worl$^{(1)}$, S.C.~Huckett$^{(1)}$,
B.I.~Swanson$^{(1)}$, A.~Saxena$^{(2)}$, \\
A.R.~Bishop$^{(2)}$, and J.~Tinka Gammel$^{(2,\rm a)}$
\\ \\
$^{(1)}$Isotope and Structural Chemistry Group, \\
$^{(2)}$Theoretical Division and \\
Centers for Nonlinear Studies and Materials Science, \\
Los Alamos National Laboratory, Los Alamos, NM 87545 USA
}

\begin{document}

\maketitle

\begin{abstract}

Resonance Raman experiments on doped and photoexcited
single crystals of mixed-halide $MX$ complexes ($M$=Pt; $X$=Cl,Br)
clearly indicate charge separation: electron polarons preferentially
locate on PtBr segments while hole polarons are trapped within PtCl
segments.  This polaron selectivity, potentially very useful for
device applications, is demonstrated theoretically using a
discrete, 3/4-filled, two-band, tight-binding, extended
Peierls-Hubbard model.  Strong hybridization of the PtCl and PtBr
electronic bands is the driving force for separation.
\end{abstract}


Halogen-bridged mixed-valence transition metal linear chain complexes
(or $MX$ chains) are highly anisotropic, quasi one-dimensional (Q1D)
materials related to conducting polymers, mixed-stack
charge-transfer salts, and oxide superconductors in terms of their low
dimensionality and competing electron-phonon ($e$-ph) and
electron-electron ($e$-$e$) interactions \cite{R1}.
A typical crystal consists of an array of
alternating metal ($M$: Pt,Pd,Ni) and halogen ($X$: Cl,Br,I) atoms;
with ligands attached to the metals, and in some cases counterions
between the chains to maintain charge neutrality.
Electrical conductivities range from values typical for
insulators to those for small-gap semiconductors.
In this class of Q1D materials,
the best studied examples of pure $MX$ materials
are based on the ethylenediamine (en) complexes,
[Pt(en)$_2$][Pt(en)$_2X_2$]$\cdot$(ClO$_4$)$_4$,
hereafter referred to as Pt$X$.
Experimentally, this large class of single crystal pure
$MX$ materials can be
systematically tuned using pressure \cite{R2}, doping \cite{R3},
and chemical variations of $M$, $X$, and the ligands \cite{R1} between
various ground state extremes; namely from
the valence-localized, strongly Peierls distorted
charge-density-wave (CDW) regime (PtCl), to the
valence-delocalized weak CDW regime (PtI), to the undistorted
spin-density-wave (SDW) phase observed in NiBr \cite{R4}.
One experimental manifestation of the tunability is absorption
spectra: the intervalence charge transfer (IVCT) absorption band
edge for PtCl, PtBr, and PtI occurs at $\sim$ 2.4, 1.5,
and 1.0 eV, respectively.  From a theoretical perspective,
strong competitions for broken-symmetry ground states
such as bond-order-wave (BOW), CDW, SDW, and possibly
spin-Peierls in these materials are governed by {\it both}
the $e$-ph and $e$-$e$ interactions, as well as dimensionality.
Furthermore,
the $MX$ chains provide us with an opportunity to probe doping- and
photo-induced local defect states \cite{R5,R6} (kinks, polarons,
bipolarons, excitons) and their interactions in controlled
environments for a wide variety of novel ground states.

In this Letter we extend our combined program on
synthesis, characterization, and modeling of the $MX$ class of
materials from pure $MX$ chains to {\it mixed-halide}
$MX_xX^\prime_{1-x}$ systems.
We find direct spectroscopic evidence
(vibrational modes and electronic absorption)
of {\it charge separation} in the mixed-halide systems.  The focus
here is the novel charge separation and its potential device
applications.

We recently reported the preparation
of a series of PtCl$_x$Br$_{1-x}$ mixed-halide materials that form
relatively defect-free, macroscopically homogeneous
crystalline solids over the entire 0$<$$x$$<$1 range due to
nearly identical crystal structure parameters \cite{R9}.
We have also prepared mixed-halide solids consisting
of PtCl$_x$I$_{1-x}$ and PtBr$_x$I$_{1-x}$, that, due to
phase separations, form only in the narrow doping range near the pure
PtCl, PtBr, or PtI compositions.  Significantly, these mixed
materials consist of segments of pure PtBr and PtCl producing
interfaces (or junctions) between the two distinct halide segments,
rather than being ``alloys'' with $X^\prime$ randomly replacing $X$.
Thus, within a single PtCl$_x$Br$_{1-x}$ crystal there
are long segments of PtCl and PtBr with spectral signatures
unperturbed from the respective pure materials.  This
offers the opportunity to study a class of materials that contain
interfaces {\it within a single crystal}.

Theoretically, we used a Hartree-Fock (HF) spatially inhomogeneous
mean-field approximation to study the electronic structure \cite{R10},
and a direct-space random phase approximation to investigate
phonons \cite{R11} (and associated ir and resonance Raman (RR) spectra)
in appropriate many-body Hamiltonians \cite{half_vs_34}.
Direct comparison between experiment and theory substantiates charge
separation in mixed-halide systems: electron polarons locate
in the PtBr segment while hole polarons locate
in the PtCl segment of a PtCl$_x$Br$_{1-x}$ chain.
Though we only report here on the PtCl/PtBr system, we find
polaronic selectivity for all Pt-based mixed-halide
systems (Cl/Br, Br/I, and Cl/I).  From a band structure
point of view, mixed-halide chains represent a 1D analog
of heterojunctions in semiconductors. In this
context, we emphasize that the charge separation is a
nontrivial result of {\it strong} lattice relaxation and not
simply undoped band structure considerations.

Synthesis and composition of pure PtCl, pure PtBr, and mixed PtCl/PtBr
solids are described elsewhere \cite{R9,R12}.
The electronic structure was probed using RR excitation profiles.
In all cases a crystal was divided and one
portion used for Raman measurements and the other for chemical
analysis.  Raman spectra were obtained from single crystals
at 13$\pm$2 K with incident intensities less than 2 mW.
The experimental procedures, instrumentation, and
RR signatures of electron ($P^-$) and
hole ($P^+$) polarons in pure $MX$ chains were reported
earlier \cite{R5,R6,R7,R8}, enabling us to address here the
distinctive new phenomena associated with mixed-halide configurations.

Upon photolysis of mixed-halide materials, photoinduced charged
defects are produced in high concentration and become preferentially
located on particular chain segments. In Figs.~\ref{F1}(a)
and \ref{F1}(b), the RR spectrum probed at 1.34 eV of a mixed
PtCl$_{.75}$Br$_{.25}$ crystal is shown before and after
photolysis at 2.54 eV (beyond the band edge of pure PtCl).
Fig.~\ref{F1}(c) illustrates the difference spectrum between
pre- and post-photolysis results.  This is an excitation
energy (1.34 eV) where resonance enhancement is known to
occur for the $P^+$ and $P^-$ in PtCl, and a region
of pre-resonance for the electron bipolaron and $P^-$ in PtBr.
As shown in Fig.~\ref{F1}(a), prior to photolysis the
Cl$-$Pt\lower2.5pt\hbox{$^{^{IV}}$}\hskip-3pt$-$Cl symmetric
stretch chain mode at 308 cm$^{-1}$ dominates the spectra.
Weak features at $\sim$285 and 325 cm$^{-1}$ are attributed
to $P^+$ defects and edge state modes in PtCl segments,
respectively.  In the spectral region for
PtBr, modes at 213 and 181 cm$^{-1}$ are observed which are
vibrational signatures for PtBr edge states.
(Our calculations yield an edge mode at $\sim$210 cm$^{-1}$
for a Cl$-$Pt\lower2.5pt\hbox{$^{^{IV}}$}\hskip-3pt$-$Br interface.)
Also observed are relatively weak features
at 196, 174, and 150 cm$^{-1}$.  The broad feature at 150 cm$^{-1}$
has been ascribed to a $P^-$ in a PtBr segment \cite{R7}.
The distinct features between 181 and 171 cm$^{-1}$ modes
are attributed to the
Br$-$Pt\lower2.5pt\hbox{$^{^{IV}}$}\hskip-3pt$-$Br symmetric stretch
for short correlation length chain segments \cite{R13}.
The characteristic chain mode observed in pure PtBr (166 cm$^{-1}$)
is not seen, indicating that long Br segments ($\geq$10 PtBr)
are not present in the material.

Photolysis within the band gap of PtCl and in the ultragap region of
PtBr (2.54 eV) causes an increase in $P^+$ defects localized
within the PtCl segments at $\sim$285 cm$^{-1}$ [Figs.~\ref{F1}(b)
and \ref{F1}(c)].
These results are in contrast to the photolysis of pure materials:
in pure crystals of PtCl, RR studies show an increase
in {\it both} $P^-$ (263 cm$^{-1}$) and $P^+$ (287 cm$^{-1}$) local
modes.  Significantly, in the mixed materials, {\it no} $P^-$
features appear in the PtCl segments with photolysis.
In the PtBr segments, the $P^-$ defect at 150 cm$^{-1}$ also
increases in intensity.  Within this region, a general increase in
intensity is observed for features from 196-174 cm$^{-1}$.
The characteristic features associated with the PtBr symmetric
stretch from different chain lengths are unresolved and complicated.
A further observation is that the edge state modes at $\sim$210
and 181 cm$^{-1}$ decrease in intensity, indicating
there is a loss of unperturbed edge state.

In a second experiment, photolysis was done on a mixed crystal of
PtCl$_{.95}$Br$_{.05}$ (Fig.~\ref{F2}).  This material exhibited
similar results within the PtCl regions: the growth of
the $P^+$ features ($\sim$285 cm$^{-1}$) occurred after photolysis.
Within the PtBr regions, one broad feature at 180 cm$^{-1}$
and one sharp feature at 194 cm$^{-1}$ increase in
intensity.  These features are consistent with the 25\% PtBr mixed
crystal, yet no features at 150 cm$^{-1}$ are observed.
Knowing these 5\% PtBr photolysis results, three features are now
apparent in the 140-200 cm$^{-1}$ region for the 25\% PtBr
crystal [Fig.~\ref{F1}(c)]: two broad features
at 150 and 180 cm$^{-1}$, and one sharp feature at 194 cm$^{-1}$.
The 150 cm$^{-1}$ feature arises from a $P^-$ within PtBr
segments, observed for these mixed crystals only at
the higher doping when PtBr
correlation lengths are fairly long ($\sim$10 Pt) \cite{R7}.
The features at 180 and 196 cm$^{-1}$
that grow in upon photolysis are due to an electron defect pinned
at a Cl$-$Pt\lower2.5pt\hbox{$^{^{IV}}$}\hskip-3pt$-$Br
edge \cite{R13}.  This creates a loss of
Cl$-$Pt\lower2.5pt\hbox{$^{^{IV}}$}\hskip-3pt$-$Br
edge character and a decrease in the 181 and 210 cm$^{-1}$
features observed in Fig.~\ref{F1}(c).

We studied the photoinduced
defects of mixed PtCl/PtBr materials with varying stoichiometries.
In  {\it all cases} we observed selective localization of hole
defects with {\it no} growth of electron features in PtCl segments.
We further have
discrete evidence for the formation of electron defects in PtBr
segments.  PtCl/PtI mixed materials also generate stable charge
defects, yet after photolysis in these systems, there is
a dramatic increase in the
$P^-$ defects in PtCl segments.  This result indicates that the
nature of charge separated state (electrons localized in a weak
CDW and holes localized in a strong CDW) can not be predicted
from the strength of CDW but must be due to other effects,
such as selective excitation energies.

Turning to our theoretical modeling, we consider an isolated
mixed-halide chain in which we replace a segment,
containing $m$ $X$ atoms, by $m$ $X^\prime$ atoms
where $X,X^\prime$=Cl,Br.  Focusing on the metal d$_{z^2}$
and halogen p$_z$ orbitals and including only the nearest
neighbor interactions we construct the following {\it two band}
tight-binding many-body Hamiltonian \cite{R10}:
\begin{eqnarray}
H &=&
\sum_{l\sigma}
\bigl(-t_0+\alpha\Delta_l\bigr)
                \bigl(c^\dagger_{l\sigma} c_{l+1\sigma}
                     +c^\dagger_{l+1\sigma} c_{l\sigma}\bigr)~
+\sum_{l\sigma}
\bigl[\epsilon_l-\beta_l (\Delta_l+\Delta_{l-1})\bigr]
               c^{\dagger}_{l\sigma} c_{l\sigma}
\nonumber\\ &&
+\sum_{l} U_l n_{l\uparrow} n_{l\downarrow}~
+ {1\over 2} \sum_l K_l\Delta_l^2~
+ {1\over 2} \sum_l K^l_{MM} (\Delta_{2l}+\Delta_{2l+1})^2~,
\label{E1}
\end{eqnarray}
\noindent
where $c^\dagger_{l\sigma}$ ($c_{l\sigma}$) denotes the creation
(annihilation) operator for the electronic orbital at the $l$th atom
with spin $\sigma$.  $M$ and $X$ (or $X^\prime$) occupy even and
odd sites, respectively.
$\Delta_l := \hat y_{l+1} - \hat y_l$, where $\hat y_l$ are the
displacements from uniform lattice spacing of the atoms at site $l$.
Eq.~(\ref{E1}) includes as parameters the on-site energy or electron
affinity $\epsilon_l$ ($\epsilon_M$$=$$e_0$, $\epsilon_X$$=$$-e_0$,
$\epsilon_{X^\prime}$$=$$e_0$$-$$2e^\prime_0$),
electron hopping $(t_0, t_{0}^\prime)$,
on-site $(\beta_M, \beta_X, \beta_{X^\prime})$ and
inter-site $(\alpha, \alpha^\prime)$ $e$-ph coupling, on-site
$e$-$e$ repulsion $(U_M, U_X, U_{X^\prime})$, and finally
effective $M$-$X$ ($K$) or $M$-$X^\prime$ ($K^\prime$) and $M$-$M$
($K_{MM}, K^{\prime}_{MM}$) springs to model the elements of
the structure not explicitly included.  In particular, $K_{MM}$
and $K^\prime_{MM}$ account for the (halide-dependent) rigidity of
the metal sublattice connected into a 3-dimensional network via
ligands \cite{interchain}.  Periodic boundary conditions were employed.
Long-range Coulomb fields have also been studied \cite{R14}.
At stoichiometry there are 6 electrons per $M_2X_2$
(or $M_2X^\prime_2$) unit, or 3/4 band filling.
Note the metal $M$=Pt energies $\epsilon_{2l}$ are the same in both
segments.  The interface (edge) between the two segments can
be of two types: (1) centered on a reduced metal
site (Pt\lower2.5pt\hbox{$^{^{II}}$\hskip-2pt,} the long-long bond)
or (2) centered on an oxidized metal site
(Pt\lower2.5pt\hbox{$^{^{IV}}$}\hskip-3pt, the short-short bond).
The corresponding electronic and phonon spectra, and thus the
associated optical and Raman spectra, differ for the two cases,
though the general features of charge separation are unaltered.
Below we consider case (1) only.
A combination of ground state experimental data
and quantum chemical and band structure
calculations have lead us to the effective parameter sets for the
Hamiltonian, Eq.~(\ref{E1}), listed in Table~1 \cite{R10}.

Fig.~\ref{F3}(a) shows a representative mixed-halide chain
considered in
our numerical simulations.  It contains 24 Pt and 24 halogens, with
a segment containing 8 Cl replaced by 8 Br. (The results are
not sensitive to segment length.)
The electronic wavefunctions and spectra  (not shown here)
indicate that the PtCl and PtBr bands are strongly hybridized.
The highest occupied level (36) and the lowest unoccupied
level (37) are PtBr-like.   Therefore, one might naively
expect doping- and photo-induced electrons
as well as holes to locate on PtBr segment.

However, the situation changes dramatically upon
doping.  The electronic spectrum in the presence of a $P^-$
is similar to the undoped chain except that the $P^-$
levels (36, 37) move into the gap, retaining their predominantly
PtBr character.  Thus a $P^-$ locates on
the PtBr segment, consistent with experimental observation.
For the electronic spectrum in the presence of a $P^+_\downarrow$,
we find the wavefunction $\psi^{36}_\uparrow$ is still Br-like
(though now unoccupied), but  $\psi^{36}_\downarrow$ (occupied) is
no longer PtBr-like.
Instead it becomes PtCl-like, as evidenced in Fig.~\ref{F3}(b).
Note the energies, $E^{36}_\uparrow$ and $E^{36}_\downarrow$,
split, as shown in Fig.~\ref{F3}(c),
with $E^{36}_\downarrow$ and $E^{37}_\uparrow$ becoming
nearly degenerate, due to the presence
of a Hubbard term ($U$) in the Hamiltonian, Eq.~(\ref{E1}).
This conversion of a PtBr-like level
into a predominantly PtCl-like level subsequent to doping is a direct
consequence of strong lattice relaxation in the mixed-halide
systems.  In other words, the strong hybridization of PtCl and PtBr
levels is nontrivially affected when a charge (polaron) is added to
the system.  Such a strong lattice relaxation effects are unusual for
conventional semiconductors and many other narrow gap materials.

To further illustrate charge separation we systematically
studied the lattice relaxation. Upon doping, if we initially
placed a $P^-$ in the PtCl segment ($i.e,$, the ``wrong" segment)
and allowed the system to self-consistently evolve to the
minimum energy configuration, we found the $P^-$  migrated
to a Pt\lower2.5pt\hbox{$^{^{IV}}$} site in the PtBr segment.
A $P^+$ placed in the PtBr segment similarly migrated
to a Pt\lower2.5pt\hbox{$^{^{II}}$} site in the PtCl
segment \cite{qualify}.
Analogously, the photogenerated exciton was invariably
unstable and broke into a $P^+$ and a $P^-$,
the $P^+$ migrating to the PtCl segment and its companion $P^-$
to the PtBr segment \cite{qualify2}.  These cases are
illustrated in Fig.~\ref{F4}. Both doping and photoexcitation
lead to charge separation with a specific
polaron charge selectivity consistent with experimental observations.

In conclusion, we have clearly demonstrated the existence of charge
separation in mixed-halide systems (with PtCl/PtBr as an example) both
experimentally and in theoretical calculations.  In particular
the doping-  and photo-induced, charged, polaronic defects
preferentially locate on
one part of the chain, say on $MX^\prime$, rather than on $MX$. This
selectivity and charge storage depend on the nature of the polaronic
excitations as well as the choice of $X$ and $X^\prime$ but not on
segment lengths.  Certain excitations  (such as kink solitons), however,
show a tendency for pinning at or in the vicinity of the
interface between $MX$ and $MX^\prime$.  Moreover, the
halogen atoms at the edge are involved in charge transfer across the
edge, and the $MX$ unit adjacent to the edge is polarized, with
distinctive  local ir and Raman ``edge'' modes.
Thus, the mixed-halide systems are of significant interest
in utilizing charge separation in potential photovoltaic
or photoconductive device applications.  To further
our understanding in this regard we are also beginning to
explore mixed-metal (e.g. PtBr/PdBr) and $MMX$ systems.

We acknowledge fruitful discussions with R.J. Donohoe and X.Z. Huang.
This work was supported by the US DOE.
JTG holds a NRC-NOSC Research Associateship.


\vfil\eject
\vbox{
{Table 1:
Parameters for the Pt$X$ materials in Eq.~(\ref{E1}).
$E_g$ is the IVCT band edge, and $\beta_X=-0.5\beta_M$.}
\hfil\break
\begin{tabular}{c|cc|cccccccc}
\hline
\hline
\smallskip
$MX$ & $\Delta$ & $E_g$
& $t_0$ & $\alpha$ & $e_0$ & $\beta_M$
& $U_M$ & $U_X$ & $K$ & $K_{MM}$ \\
& (\AA) & (eV)
& (eV) & (eV/\AA) & (eV) & (eV/\AA)
& (eV) & (eV) & (eV/\AA$^2$) & (eV/\AA$^2$) \\
\hline
PtCl &0.38  &2.50 &1.02  &0.5 &2.12 &2.7  & 1.9 & 1.3  &6.800 & 0.0 \\
PtBr &0.24  &1.56 &1.26  &0.7 &0.60 &1.8  & 0.0 & 0.0  &6.125 & 0.7 \\
\hline
\hline
\end{tabular}
}

\vfil\eject
\centerline{FIGURES}

\figure{\label{F1}
Resonance Raman (RR) spectra at 1.34 eV of PtCl$_{.75}$Br$_{.25}$
mixed-halide crystal (a) before and (b) after photolysis with 2.54 eV
excitation at 2 mW for 1.5 hours at 13 K. (c) Difference spectra:
Post-photolysis intensity minus pre-photolysis intensity.
}

\figure{\label{F2}
Difference RR spectra at 1.34 eV of
PtCl$_{.95}$Br$_{.05}$ mixed-halide crystal under the experimental
conditions of Fig.~\ref{F1}.
}

\figure{\label{F3}
(a) A PtCl chain containing a segment of PtBr.  Note that both edges
of the segment are Pt\lower2.5pt\hbox{$^{^{II}}$} sites.
(b) Electronic up (-\,-\,-)  and down (---)  spin wavefunctions
at the Fermi surface (level 36).
(c) Localized electronic levels for $P^+_\downarrow$.
}

\figure{\label{F4}
Excess charge (---) and spin (-\,-\,-) densities
with respect to the mixed-chain ground state
showing polaron migration for (a) $P^-_\uparrow$ and
(b) $P^+_\downarrow$ on a 16 PtCl and 8 PtBr unit chain, and
(c) subsequent to photoexcitation of an up spin electron
between levels 36 and 37 on an 8 PtCl and 16 PtBr unit chain.
The large circles indicate initial defect
location, though the result is insensitive
to initial condition \cite{qualify,qualify2}.
}


\begin{thebibliography}{10}
\bibitem{half_vs_34}
In contrast to 1/2-filled bands, comparison of HF and exact
results indicates that for 3/4-filled systems the HF results
are qualitatively reliable (and quantitatively for PtBr).
Also, for the heavy atoms
considered here, quantum lattice and disorder effects are minimal.
(See Ref.~\cite{R10}).
\bibitem{interchain}
Our theoretical and experimental studies indicate that interchain
interactions are not important in $MX$ systems. However, we intend to
quantify interchain effects in future studies.
\bibitem{qualify}
Local energy minima are occasionally found where
the $P^+$ localizes in the PtBr segment.
The presence and location of such metastable configurations
depend on the length of both segments.
\bibitem{qualify2}
No metastable minima such as bound $P^+ - P^-$ pairs or
excitons are found in photoexcitation since $P^-$ and $P^+$ repel due
to (a) strong lattice relaxation, and (b) different on-site energies
in the $MX$ and $MX^\prime$ regions. This is in contrast to the pure
materials, where the question of exciton creation is more subtle.
\end{thebibliography}
\end{document}